\documentclass[12pt]{article}
\usepackage{graphicx} 
\usepackage{epstopdf}
\usepackage{amsthm,amsmath,amssymb, amsfonts, amscd, xspace, pifont}
\usepackage{epsfig, psfrag,pstricks}
\usepackage{times}
\usepackage[authoryear,round]{natbib}
\usepackage{algorithm, algorithmic}
\usepackage{booktabs}
\usepackage{multirow}
\usepackage{makecell}
\usepackage{array}
\newtheorem{theorem}{Theorem}

\providecommand{\keywords}[1]{\textbf{\textit{Keywords: }} #1}

\def\btheta{\mbox{\boldmath $\theta$}}
\def\bepsilon{\mbox{\boldmath $\epsilon$}}

\def\bSigma{\mbox{\boldmath $\Sigma$}}

\def\bw{\mbox{\boldmath $\beta$}}

\def\bR{\mathbb{R}}
\def\mE{\mathbb{E}}
\def\bP{\mathbb{P}}
\def\mQ{\mathbb{Q}}

\def\bmu{\mbox{\boldmath $\mu$}}
\def\ba{{\bf a}}
\def\bb{{\bf b}}

\def\bm{{\bf m}}

\def\bw{{\bf w}}

\def\bu{{\bf u}}

\def\bI{{\bf I}}

\def\bx{{\bf x}}
\def\by{{\bf y}}

\def\nn{\nonumber}

\def\v2{\vspace{0.2in}}

\numberwithin{equation}{section}
\begin{document}
\setcounter{page}{1}
\baselineskip=15pt
\footskip=.3in

\title{One-step data-driven generative model via Schr\"odinger Bridge}

\date{}

\author{Hanwen Huang \\\\
    {\it Department of Biostatistics, Data Science and Epidemiology}\\
    {\it Medical College of Georgia}\\
    {\it Augusta University, Augusta, GA 30912}}

\maketitle

\begin{abstract}
Generating samples from a probability distribution is a fundamental task in machine learning and statistics. This article proposes a novel scheme for sampling from a distribution for which the probability density $\mu(\bx)$ for $\bx\in\bR^d$ is unknown, but finite independent samples are given. We focus on constructing a Schr\"odinger Bridge (SB) diffusion process on finite horizon $t\in[0,1]$ which induces a probability evolution starting from a fixed point at $t=0$ and ending with the desired target distribution $\mu(\bx)$ at $t=1$. The diffusion process is characterized by a stochastic differential equation whose drift function can be solely estimated from data samples through a simple one-step procedure. Compared to the classical iterative schemes developed for the SB problem,  the methodology of this article is quite simple, efficient, and computationally inexpensive as it does not require the training of neural network and thus circumvents many of the challenges in building the network architecture. The performance of our new generative model is evaluated through a series of numerical experiments on multi-modal low-dimensional simulated data and high-dimensional benchmark image data. Experimental results indicate that the synthetic samples generated from our SB Bridge based algorithm are comparable with the samples generated from the state-of-the-art methods in the field. Our formulation opens up new opportunities for developing efficient diffusion models that can be directly applied to large scale real-world data.  
\end{abstract}
 
\keywords{Diffusion process; Generative model; Neural network; Sampling; Stochastic differential equation.}

\section{Introduction}
Generating samples efficiently from complex probability distributions plays a key role in a variety of prediction and inference tasks in machine learning and statistics. We have two types of settings. In the first setting, the probability distribution $\mu(\bx)$ for $\bx\in\bR^n$ is explicitly given. This is often the case in Bayesian statistics. In the second setting, the explicit form of $\mu(\bx)$ is not specified, but we have access to a collection of i.i.d. random samples. This is the case in generative modeling which has been extensively studied over the last years and has found wide-ranging applications across diverse domains such as computer vision \citep{huang2021multimodal,sohldickstein2015deep,Endo_2020}, image analysis \citep{wang2021geometry,pu2015deep}, natural language processing \citep{inproceedings,yao2019DGM4NLP,Miao2017DeepGM}, drug discovery \citep{16272714120230328,ZENG2022100794}, and recommendation systems \citep{JMLR:v22:20-1098}. Most of the existing generative models seek to understand the underlying data generating mechanisms through learning a nonlinear function that can transform a simple reference distribution to the target distribution. Here we focus on a powerful class of generative models, called the diffusion models which have achieved enormous success in synthesizing high-fidelity data. Instead of assume a specific form of the data distribution, the diffusion models implement a stochastic process and use its endpoint to represent the target distribution. Two promising approaches in diffusion models are score based diffusion models \citep{sohldickstein2015deep,ho2020denoising,song2021maximum,song2021scorebased} and Schr\"odinger Bridge based diffusion models  \citep{bernton2019schrodinger,chen2023likelihood,debortoli2023diffusion,shi2023diffusion,chen2023schrodinger,liu2023schrodinger,pmlr-v206-stromme23a}. 

The score based generative models (SGM) consist of two processes: the noise-injecting process also called the forward process and the denoising process also called the backward process. The forward process is implemented by incrementally adding noise to the original data until they are indistinguishable from samples drawn from a easily sampled prior distribution, e.g., the Gaussian distribution. The backward process is implemented by recovering the original data distributions from the prior. Both processes can by modeled as stochastic processes in terms of a stochastic differential equation (SDE). The drift function for the forward SDE is defined such that the desired prior becomes the marginal distribution of the process at the end time. Similarly, the drift function for the backward SDE is defined such that it can be used to generate data samples from the prior. The basis of SGM lies in the notion of the score function (i.e. the gradient of the log-density function of a given distribution) which can be learned (often parametrized by neural network) through regression at each time step. 

Despite its empirical success, the diffusion process of SGM needs to run sufficiently large time steps so that the end distribution is approximately Gaussian. To overcome this limitation, an alternative approach based on the so-called Schr\"{o}dinger Bridges (SB) has been proposed which allows one to obtain exact samples from a sufficiently well-behaved target density in finite time. This approach, going back to a problem posed by Schrödinger in the 1930s in the context of physics \citep{schrodinger1932theorie,schrodinger1931umkehrung}, tries to construct a stochastic process that has two given distributions as marginals at the initial and end times, while subject to a reference process. SB can also be formulated as an entropically regularized optimal transport problems which seeks optimal policies to transform the probability density between given arbitrary initial and end distributions in a finite horizon. Despite its attractiveness, the numerical solution of SB problems are usually based on the so-called Iterative Proportional Fitting (IPF) algorithm (also known as the Sinkhorn algorithm) which is computationally more demanding in contrast to the SGM methods \citep{ChenOptimal}. IPF solves the bridge problem by creating a convergent sequence of forward and backward processes, known as half-bridges, in which only one of the two distributions at the boundaries of the time interval is kept fixed. The drift function for the corresponding SDE of a half-bridge is learned from samples created by the half-bridge of the previous iteration using a regression approach via training a deep neural network. This approach is deemed as computationally too demanding because the drift at a given iteration is represented as a sum of scores obtained in the previous steps. Moreover, it requires the storage of an increasing number of neural networks, two per iteration. Consequently, practical usages of SB for learning high-dimensional data distributions is limited for these reasons. 

In this work, we propose an efficient method to generate samples from a distribution for which the probability density is unknown, but a finite set of sample points are provided. Our approach is based on Euler-Maruyama discretization of a tractable class of SB diffusion processes, defined on the unit time interval $[0,1]$, that transport the degenerate Dirac delta distribution at time zero to the target distribution at time one. We show that, given such boundary pairs, the solutions of the SB system admit analytic forms which solely rely on the given data samples and thereby yield a simulation-free framework that avoids unfavorable complexity occurring in many simulation-base algorithms. Our approach differs from prior SB based work in that it is a one-step solution which does not need to run iteration. Moreover, our method scales well to high-dimensions  as it does not require training neural network and thus substantially reduces the computation burden. We demonstrate the practical use of our algorithm on simulated and real-world data by quantitatively and qualitatively comparing its performance to the state-of-the-art methods in the field. 

A fundamental difference between our SB based approach and many existing related works is that we start the process from a Dirac delta distribution instead of a Gaussian distribution. For Gaussian and target boundary pair, usually the SB problem has no closed form solution and one has to rely on iterative schemes. In addition to IPF, various other type of iterative algorithms have been proposed in the literature. Examples include \cite{bernton2019schrodinger,e23091134,pavon2018datadriven,chen2023likelihood,e25020316,debortoli2023diffusion,e25020316,ChenOptimal,shi2023diffusion,chen2023schrodinger,liu2023schrodinger,Lee_Lee_Bang_Kim_2024} and many others. Recently, the Schr\"{o}dinger-F\"{o}llmer process (SFP) has been used for developing sampling scheme which considers the same boundary pair and transports the delta distribution at time zero to the target distribution at time one. For example, the SFP based algorithms are proposed in \cite{zhang2022path,huang2021schrodingerfollmer} for drawing samples from the target density which is known up to a normalizing constant. This approach is also applied to the problem of Bayesian inference in \cite{vargas2022bayesian}. Our method is different from these SF based approaches in that we are considering sampling scheme from distributions which are known only through a finite set of sample points. \cite{wang2021deep} present a a two-stage deep generative model based on SF, in which the first stage learns a smoothed version of the target distribution, and the second stage derives the actual target at the sample level. Our method can be considered as a one-step simplification of this approach and therefore is more computationally simple. Most closely related to the current paper are results by \cite{hamdouche2023generative} that derives a generative model for time series based on SFP approach in which the drift function is directly estimated from the data samples. The major difference is that \cite{hamdouche2023generative} focuses on the time series problem, while we focus on the problem of drawing samples from arbitrary probability distributions. In contrast to the above SFP based algorithms that only consider standard Brownian motion as reference process, our approach is more flexible and allows the reference process to be any Ito's process. It is shown in the numerical studies in Section \ref{numerical} that the performance of our algorithm in some situations can be significantly improved by appropriately choosing the reference process. 

The paper is organized as follows. In Section \ref{method}, we provide an introduction to Schr\"{o}dinger bridge diffusion process and present a class of tractable Schr\"{o}dinger system of which the solution can be explicitly derived. Then we propose a one-step data-driven generative model and outline its implementation algorithm. Section \ref{numerical} is devoted to numerical studies. We test the performance of our algorithm on both simulated low-dimensional data and real-world high-dimensional image data. Concluding remarks are given in Section \ref{conclusion}. Proof for the theorem is provided in Appendix \ref{proof}. 

\section{Data-driven Schr\"odinger Bridge sampler}\label{method}

In this section we first provide some background on the Schr\"{o}dinger Bridge process. Then we introduce a class of tractable Schr\"{o}dinger Bridge processes. Finally we propose a novel generative model based on this tractable class by appropriately choosing the reference SDEs.

\subsection{Background on Schr\"{o}dinger Bridge}
We are interested in generating synthetic data from an unknown distribution $\mu$ for which we have a data set of i.i.d. samples $\bx^{(i)}\in\bR^d$, $i=1,\cdots,n$. Our algorithms are implemented through constructing a stochastic diffusion process $\{\bx_t\}_{t=0}^1$ indexed by a continuous time variable $t\in[0,1]$ such that $\bx_0\sim\delta_{\ba}$, the Dirac delta distribution centered at $\ba\in\bR^d$, and $\bx_1\sim\mu$, the target distribution. Denote by $\Omega=C([0,1],\bR^d)$ the space consisting of all $\bR^d$-valued continuous functions on the time interval $[0,1]$ and $\bP$ the measure over $\Omega$ induced by the following SDE
\begin{eqnarray}\label{sde0}
d\bx_t=\bb(\bx_t,t) dt+\sigma(t) d\bw_t,~\bx_0=\ba,
\end{eqnarray}
where $\bw_t$ is the standard $d$-dimensional Brownian motion, $\bb(\cdot,\cdot):\bR^d\times[0,1]\rightarrow\bR^d$ is a vector-valued function called the drift coefficient of $\bx_t$, and $\sigma(t):\bR\rightarrow\bR$ is a scalar function known as the diffusion coefficient of $\bx_t$. When $\bb(\bx,t)=0$ and $\sigma(t)=1$, $\bx_t$ is just the standard $d$-dimensional Brownian motion. The SDE (\ref{sde0}) has a unique strong solution as long as the drift coefficient function $\bb(\bx,t)$ is globally Lipschitz in both state $\bx$ and time $t$ \citep{10.5555/129416}. 

Denote by $\mQ_t$ the marginal probability law at time $t$ for the probability measure $\mQ$ on $\Omega$. We write ${\cal D}(\delta_{\ba},\mu)=\{\mQ:\mQ_0=\delta_{\ba},\mQ_1=\mu\}$ for the set of all path measures with given marginal distribution $\delta_{\ba}$ at the initial time and $\mu$ at the final time. Then the solution of the Schr\"{o}dinger Bridge (SB) problem with respect to the reference measure $\bP$ can be formulated by the following optimization problem
\begin{eqnarray}\label{kld0}
\mQ^\star&=&\text{argmin}_{\mQ\in{\cal D}(\delta_{\ba},\mu)}D(\mQ\|\bP),
\end{eqnarray}
where $D(\mQ\|\bP)$ denotes the relative entropy between two probability measures on $\mQ$ and $\bP$ which is defined as
\begin{eqnarray}\label{kldef}
D(\mQ\|\bP)&=&\left\{\begin{array}{cc}\int\log(d\mQ/d\bP)d\mQ&if~\mQ\ll\bP\\
\infty&otherwise\end{array}\right.,
\end{eqnarray}
where $\mQ\ll\bP$ denotes that $\mQ$ is absolutely continuous w.r.t. $\bP$ and $d\mQ/d\bP$ represents the Radon-Nikodym derivative of $\mQ$ w.r.t. $\bP$ 

\subsection{A tractable class of Schr\"{o}dinger Bridge}
It is known \citep{Pavon1989,DaiPra1991,Leonard2014} that if the reference process $\bP$ is induced by (\ref{sde0}),  then $\mQ^\star$ is induced by a SDE with a modified drift:
\begin{eqnarray}\label{sde}
d\bx_t=[\bb(\bx_t,t)+\bu^\star(\bx_t,t)] dt+\sigma(t) d\bw_t,~\bx_0=\ba.
\end{eqnarray}
When $\bb(\bx_t,t)=0$ and $\sigma(t)=1$, i.e. $\bP$ is a Wiener measure, the SB problem (\ref{kld0}) with initial margin $\delta_{\ba}$ is also called the Schr\"{o}dinger-F\"{o}llmer process (SFP). The properties of SFP have been explored in \cite{DaiPra1991,Leonard2014,TzenR19,huang2021schrodingerfollmer,vargas2022bayesian,zhang2022path} via closed form formulation of the drift $\bu^\star(\bx_t,t)$ in terms of expectations over Gaussian random variables known as the heat semigroup. These work also explores how the formulation of the SFP can be used in practice to construct exact sampling schemes for target distributions. Particularly, the work by \cite{huang2021schrodingerfollmer} investigates how to estimate the SFP drift in practice via the heat semigroup formulation using a Monte Carlo approximation. \cite{vargas2022bayesian} applies the SFP drift estimation to the problem of high-dimensional Bayesian inference. \cite{zhang2022path} proposes a similar algorithm to estimate the normalizing constant of un-normalized densities.

Most of the prior studies on how to use SFP to generate samples are analytical, taking as input the pre-specified probability densities. Our approach in this work, instead, inspired by the original idea of Schr\"{o}dinger, applies to situations where only samples are available. Moreover, the reference measure $\bP$ in our proposal is not restricted to the Wiener process, i.e. $\bu(\bx,t)$ and $\sigma(t)$ in (\ref{sde0}) can be any function as long as they yield a closed form solution for $\bP$. 

Denote the gradient of a smooth function  $f(\bx)$ by $\nabla f(\bx)$ and the partial derivative with respect to $\bx$ of $\psi(\bx,\by)$ for $(\bx, \by) \in \mathbb{R}^d\times\mathbb{R}^d$ by $\nabla_{\bx} \psi(\bx,\by)$. To facilitate the reformulation of the problem so that it involves the target distribution only through its available samples, we state in the following theorem the solution of the SB problem (\ref{kld0}) whose reference measure $\bP$ is induced by the SDE (\ref{sde0}) with $\bb(\bx,t)\ne 0$ and/or $\sigma(t)\ne 1$. 
\begin{theorem}\label{thm}
Assume that $\sigma(t)\in C^1([0,1])$ and the components of $\bb(\bx,t)$ are bounded continuous and satisfy Holder conditions with respect to $\bx$, i.e. there are real constants $C\ge 0,~\alpha\textgreater 0$ such that $|b_i(\bx,t)-b_i(\by,t)|\le C\|\bx-\by\|^\alpha$ for all $i=1,\cdots,d$ and $\bx,\by\in\bR^d$. Further assume that $\mu$ is absolutely continuous. Then the stochastic process (\ref{sde}) induces a probability measures $\mQ^\star$ which solves the Schrodinger Bridge problem (\ref{kld0}) if the drift term $\bu^\star(\bx_t,t)$ is given by
\begin{eqnarray}\label{drift0}
\bu^\star(\bx,t)&=&\frac{\sigma(t)^2\int\nabla_{\bx}g_t(\bx,\bx_1)\mu(d\bx_1)}{\int g_t(\bx,\bx_1)\mu(d\bx_1)},
\end{eqnarray}
where
\begin{eqnarray}\label{transition}
g_t(\bx,\bx_1)&=&\frac{q(t,\bx,1,\bx_1)}{q(0,\ba,1,\bx_1)},
\end{eqnarray}
with $q(t_1,\bx,t_2,\by)$ denotes the transition density of $\bx_{t_2}=\by$ at time $t_2$ given $\bx_{t_1}=\bx$ at time $t_1$ for stochastic process $\bx_t$ governed by the reference SDE (\ref{sde0}).
\end{theorem}
Theorem \ref{thm} shows that we can start from any point $\bx_0 = \ba$ and update the values of $\{\bx_t : 0 < t \le 1\}$ according to the SDE (\ref{sde}) in continuous time, then the value $\bx_1$ has the desired distributional property, that is, $\bx_1 \sim \mu$. Note that the drift function (\ref{drift0}) is also the solution of the following stochastic control problem
\begin{eqnarray}\label{optimal}
\bu^\star(\bx_t,t)&=&\text{argmin}_{\bu(\bx_t,t)}E\left[\frac{1}{2}\int\|\bu(\bx_t,t)\|^2dt\right],\\\nn
s.t.&&\left\{\begin{array}{c}d\bx_t=[\bb(\bx_t,t)+\bu^\star(\bx_t,t)] dt+\sigma(t) d\bw_t,\\\bx_0=\delta_{\ba},~~~\bx_1\sim\mu.\end{array}\right.
\end{eqnarray}
A nice property of the stochastic control framework (\ref{optimal}) is that it enables us to design efficient sampling scheme via transporting particles on any fixed single point $\ba$ to the particles drawn from the target distribution $\mu$ on the unit time interval \citep{zhang2022path}. 

\subsection{Drift estimation}

The drift term (\ref{drift0}) involves the integration over the target distribution $\mu$ which usually has no closed form solutions. We are considering the frequently occurring situation where $\mu$ is only known through a fixed set of $n$ samples. In order to evaluate the integration in (\ref{drift0}) accurately, we need to find a good estimation for the probability density $\hat{\mu}$ from the given samples. This is quite challenging especially in high-dimensional cases because the underlying distribution $\mu$ often has multi-modes or lies on a low-dimensional manifold, which cause difficulty to learn from simple distribution such as Gaussian or mixture of Gaussian. Many methods have been proposed to form useful estimates $\hat{\mu}$ of $\mu$ typical of which are the kernel method \citep{10.1214/aoms/1177704472},  the interpolation method \citep{10.1214/aos/1176342998}, and diffusion based method \citep{1056736}. These traditional methods are mainly designed for low-dimensional cases and fall short in scaling to high-dimensions. \cite{wang2021deep} obtained the estimator of probability density via minimizing an empirical logistic regression loss determined by the score function that can be modeled using a deep network. However, this method is very costly and prone to high variance as it requires the training of the neural network. 

In this work, we instead take two advantages of formula (\ref{drift0}) and propose an alternative one-step estimation for the drift coefficient $\bu^\star(\bx,t)$. The first advantage is that it involves expectations under the target distribution $\mu$,  so its integral can be replaced by its empirical counterpart, i.e. 
\begin{eqnarray}\label{clt}
\int g_t(\bx,\bx_1)\mu(d\bx_1)\rightarrow\frac{1}{n}\sum_{i=1}^ng_t(\bx,\bx^{(i)}), 
\end{eqnarray}
which leads to the following direct estimation
\begin{eqnarray}\label{drift}
\hat{\bu}^\star(\bx,t)&=&\frac{\sigma(t)^2\sum_{i=1}^n\nabla_{\bx} g_t(\bx,\bx^{(i)})}{\sum_{i=1}^ng_t(\bx,\bx^{(i)})}.
\end{eqnarray}
This one-step solution does not require the training of deep neural network and avoids the complexity of building the network architecture and thus is computationally more inexpensive. The second advantage is that the transition density $q(t_1,\bx,t_2,\by)$ in (\ref{transition}) is for the reference process which can be obtained in closed-forms by appropriately choosing the drift function $\bb(\bx,t)$ and diffusion function $\sigma(t)$ in (\ref{sde0}). For example, when $\bb(\bx,t)$ is affine in terms of $\bx$, the transition kernel is always a Gaussian distribution denoted by $N(\bx;\bmu,\bSigma)$, where the mean $\bmu$ and covariance matrix $\bSigma$ are often known and can be obtained with standard techniques. 

\subsection{Reference SDEs}\label{refsde}
In numerical studies, we consider three choices of reference SDEs proposed in \cite{song2021scorebased} which have been successfully used in many probabilistic generative tasks. All of them can yield closed form solutions for both the transition probability $q(s,\bx_s,t,\bx_t)$ and the drift $\bu^\star(\bx,t)$.
\begin{itemize}
\item Variance exploding (VE) SDE:
\begin{eqnarray}\nn
d\bx_t&=&\sqrt{\alpha^\prime(t)} d\bw_t,\\\nn
q(s,\bx_s,t,\bx_t)&=&N(\bx_t;\bx_s,[\alpha(t)-\alpha(s)]\bI_d),\\\label{ve}
\bu^\star(\bx,t)&=&\frac{\alpha^\prime(t){\displaystyle\int}(\bx_1-\bx)f_t(\bx,\bx_1)\mu(d\bx_1)}{(\alpha(1)-\alpha(t)){\displaystyle\int}f_t(\bx,\bx_1)\mu(d\bx_1)},
\end{eqnarray}
where $\alpha^\prime(t)=\frac{d\alpha(t)}{dt}$, $\bI_d$ is a $d$-dimensional identity matrix and
\begin{eqnarray}\label{vef}
f_t(\bx,\bx_1)&=&\exp\left(\frac{\|\bx_1-\ba\|^2}{2(\alpha(1)-\alpha(0))}-\frac{\|\bx_1-\bx\|^2}{2(\alpha(1)-\alpha(t))}\right),
\end{eqnarray}
where $\|\cdot\|$ denotes the $L_2$-norm. If we choose $\alpha(t)=t$, it is just the standard $d$-dimensional Brownian motion. 
\item Variance preserving (VP) SDE:
\begin{eqnarray}\nn
d\bx_t&=&-\frac{1}{2}\beta(t)\bx_t dt+\sqrt{\beta(t)} d\bw_t,\\\nn
q(s,\bx_s,t,\bx_t)&=&N(\bx_t;\bx_s e^{-\frac{1}{2}\int_s^t\beta(s^\prime)ds^\prime},[1-e^{-\int_s^t\beta(s^\prime)ds^\prime}]\bI_d),\\\label{vp}
\bu^\star(\bx,t)&=&\frac{\beta(t)e^{-\frac{1}{2}\int_t^1\beta(s^\prime)ds^\prime}{\displaystyle\int}\left[\bx_1-\bx e^{-\frac{1}{2}\int_t^1\beta(s^\prime)ds^\prime}\right]f_t(\bx,\bx_1)\mu(d\bx_1)}{(1-e^{-\int_t^1\beta(s^\prime)ds^\prime}){\displaystyle\int}f_t(\bx,\bx_1)\mu(d\bx_1)},
\end{eqnarray}
where
\begin{eqnarray}\nn
f_t(\bx,\bx_1)&=&\exp\left(\frac{\|\bx_1-\ba e^{-\frac{1}{2}\int_0^1\beta(s^\prime)ds^\prime}\|^2}{2(1-e^{-\int_0^1\beta(s^\prime)ds^\prime})}-\frac{\|\bx_1-\bx e^{-\frac{1}{2}\int_t^1\beta(s^\prime)ds^\prime}\|^2}{2(1-e^{-\int_t^1\beta(s^\prime)ds^\prime})}\right).
\end{eqnarray}
\item Sub-Variance preserving (sub-VP) SDE:
\begin{eqnarray}\nn
d\bx_t&=&-\frac{1}{2}\beta(t)\bx_t dt+\sqrt{\beta(t)\left(1-e^{-2\int_0^t\beta(s)ds}\right)} d\bw_t,\\\label{sub-VP}
q(s,\bx_s,t,\bx_t)&=&N(\bx_t;\bx_s e^{-\frac{1}{2}\int_s^t\beta(s^\prime)ds^\prime},[1-e^{-\int_s^t\beta(s^\prime)ds^\prime}]^2\bI_d),\\\nn
\bu^\star(\bx,t)&=&\frac{\beta(t)\left(1-e^{-2\int_0^t\beta(s^\prime)ds^\prime}\right)e^{-\frac{1}{2}\int_t^1\beta(s^\prime)ds^\prime}{\displaystyle\int}\left[\bx_1-\bx e^{-\frac{1}{2}\int_t^1\beta(s^\prime)ds^\prime}\right]f_t(\bx,\bx_1)\mu(d\bx_1)}{(1-e^{-\int_t^1\beta(s^\prime)ds^\prime})^2{\displaystyle\int}f_t(\bx,\bx_1)\mu(d\bx_1)},
\end{eqnarray}
where
\begin{eqnarray}\nn
f_t(\bx,\bx_1)&=&\exp\left(\frac{\|\bx_1-\ba e^{-\frac{1}{2}\int_0^1\beta(s^\prime)ds^\prime}\|^2}{2(1-e^{-\int_0^1\beta(s^\prime)ds^\prime})^2}--\frac{\|\bx_1-\bx e^{-\frac{1}{2}\int_t^1\beta(s^\prime)ds^\prime}\|^2}{2(1-e^{-\int_t^1\beta(s^\prime)ds^\prime})^2}\right).
\end{eqnarray}
\end{itemize}
The VE SDE yields a process with exploding variance when $t\rightarrow\infty$. In contrast, the VP SDE yields a process with bounded variance. In addition, the VP process has a constant unit variance for all $t\in[0,\infty)$ when the initial distribution of $\bx_0$ has a unit variance. The two successful classes of score-based generative models: score matching with Langevin dynamics (SMLD) and denoising diffusion probabilistic modeling (DDPM) are discretization of VE SDE and VP SDE respectively \citep{song2021scorebased}. In SMLD, one has $\alpha(t)=\sigma_{min}^2\left(\frac{\sigma_{max}}{\sigma_{min}}\right)^{2t}$ for $t\in[0,1]$ which leads to a diffusion coefficient $\sigma_{min}\left(\frac{\sigma_{max}}{\sigma_{min}}\right)^t\sqrt{2\log\frac{\sigma_{max}}{\sigma_{min}}}$. In DDPM, one has $\beta(t)=\bar{\beta}_{min}+t(\bar{\beta}_{max}-\bar{\beta}_{min})$ which leads to a diffusion coefficient $\sqrt{\bar{\beta}_{min}+t(\bar{\beta}_{max}-\bar{\beta}_{min})}$.  The detailed choice of $\alpha(t)$ and $\beta(t)$ in our algorithm is discussed in Section \ref{numset}.  It is worth mentioning that the noise level is annealed up in both SMLD and DDPM while it is annealed down in our implementation as our SB process corresponds to the reverse-time process of SMLD and DDPM.

Note that even if $\mu$ is known, the closed form expression for the drift term $\bu^\star(\bx,t)$ can only be obtained in some special situations. For example, if $\mu$ is Gaussian mixture, \cite{huang2021schrodingerfollmer} derived the explicit expression of the drift terms for VE SDE with $\sigma(t)=t$. We derive the corresponding results for VP SDE with general $\beta(t)$ in Section \ref{gmm}.

\subsection{Euler-Maruyama discretization}

Once the drift function $\bu^\star(\bx,t)$ is determined, the diffusion process (\ref{sde}) can be used to sample from the target distribution $\mu$ by transporting the initial degenerate distribution $\delta_{\ba}$ at $t = 0$ to the target $\mu$ at $t = 1$. The fact that the SDE (\ref{sde}) is defined on the finite time interval $[0, 1]$ allows us to numerically implement this approach simply through Euler-Maruyama discretization. We first take $N\ge 2$ grid points on $[0, 1]$ with $0=t_0\textless t_1\textless\cdots\textless t_N=1$ and $\delta_j=t_{j+1}-t_j$ being the step size for the $j$-th interval. Then the resulting discretization of the diffusion process (\ref{sde}) is
\begin{eqnarray}\label{euler}
\bx_{t_{j+1}}&=&\bx_{t_{j}}+\delta_j[\bb(\bx_{t_j},t_j)+\bu^\star(\bx_{t_j},t_j)]+\sigma(t_j)\sqrt{\delta_{j}}\bepsilon_{j},~j=0,1,\cdots,N-1,
\end{eqnarray}
where $\{\bepsilon_j\}_{j=1}^N$ are independent and identically distributed random vectors from $N(0,\bI_d)$. If the reference process is induced by VE SDE, the drift term $\bu^\star(\bx_{t_j},t_j)$ can be estimated as
\begin{eqnarray}\label{vedrift}
\hat{\bu}^\star(\bx_{t_j},t_j)&=&\frac{\alpha^\prime(t_j)\sum_{i=1}^n (\bx^{(i)}-\bx_{t_j})f_{t_j}(\bx_{t_j},\bx^{(i)})}{(\alpha(1)-\alpha(t_j))\sum_{i=1}^nf_{t_j}(\bx_{t_j},\bx^{(i)})},
\end{eqnarray}
where $f_{t_j}(\bx_{t_j},\bx^{(i)})$ is defined in (\ref{vef}). The drift terms for other reference SDEs can be estimated in a similar way. Based on (\ref{euler}), we can start from $\bx_{t_0}=\ba$ and iteratively update this initial value to obtain a realization of the random sample $\bx_{t_N}$ which is approximately distributed as the target distribution $\mu$ under suitable conditions. For convenience, we shall refer to the proposed sampling method as the data-driven Schr\"{o}dinger Bridge sampler (DSBS). The pseudocode for implementing DSBS is presented in Algorithm \ref{alg}. 
\begin{algorithm}[H]
	\caption{DSBS for given i.i.d. samples}
    \label{alg}
	\begin{algorithmic}[1]
\STATE Input: data set $\{\bx^{(i)}\}_{i=1}^n$, grid points $0=t_0\textless t_1\textless\cdots\textless t_N=1$ on time interval $[0,1]$ with step size $\delta_j=t_{j+1}-t_j$, starting point $\bx_{t_0}=\ba$.
\FOR{$j= 0,1,\ldots, N-1$ }
\STATE Sample random vector $\bepsilon_{j}\sim N(0,\bI_{d})$,
\STATE Compute $\hat{\bu}^\star(\bx_{t_j},t_j)$ according to \eqref{vedrift},
\STATE Update $\bx_{t_{j+1}}$ according to \eqref{euler}.
\ENDFOR
\STATE Output:  $\{\bx_{t_j}\}_{j=1}^{N}$
\end{algorithmic}
\end{algorithm}

Note that, in contrast to Langevin diffusion based sampling approaches which rely on the diffusions that reach the target distribution as their equilibrium state when time goes to infinity, the SB based dynamics are controlled and the target distribution is reached in finite time. An advantage of our DSBS sampling scheme over the traditional MCMC based methods is that ergodicity is not required \citep{huang2021schrodingerfollmer}. This is also due to the basic property of the SB diffusion (\ref{sde}) which transports the initial distribution $\delta_{\ba}$ at $t = 0$ to the exact target distribution $\mu$ at $t=1$. The sampling error of DSBS is entirely due to the approximation of the drift term via (\ref{vedrift}) and the Euler-Maruyama discretization via (\ref{euler}). The first type of approximation errors can be well controlled by the large sample theorem according to (\ref{clt}). The second type of approximation errors can also be made arbitrarily small under suitable conditions as shown in \cite{huang2021schrodingerfollmer}. 

\section{Numerical experiments}\label{numerical}

In this section, we demonstrate the effectiveness of our algorithm on several simulated and real data sets. We first employ two-dimensional toy examples to show the ability of our algorithm to learn multimodal
distributions. Next, we show that our algorithm is able to generate realistic image samples. We use two benchmark datasets including MNIST \citep{lecun-mnisthandwrittendigit-2010} and CIFAR-10 \citep{Krizhevsky2009LearningML}.

\subsection{Setup}\label{numset}
\begin{figure}[hbtp]
	\begin{center}
		\epsfig{file=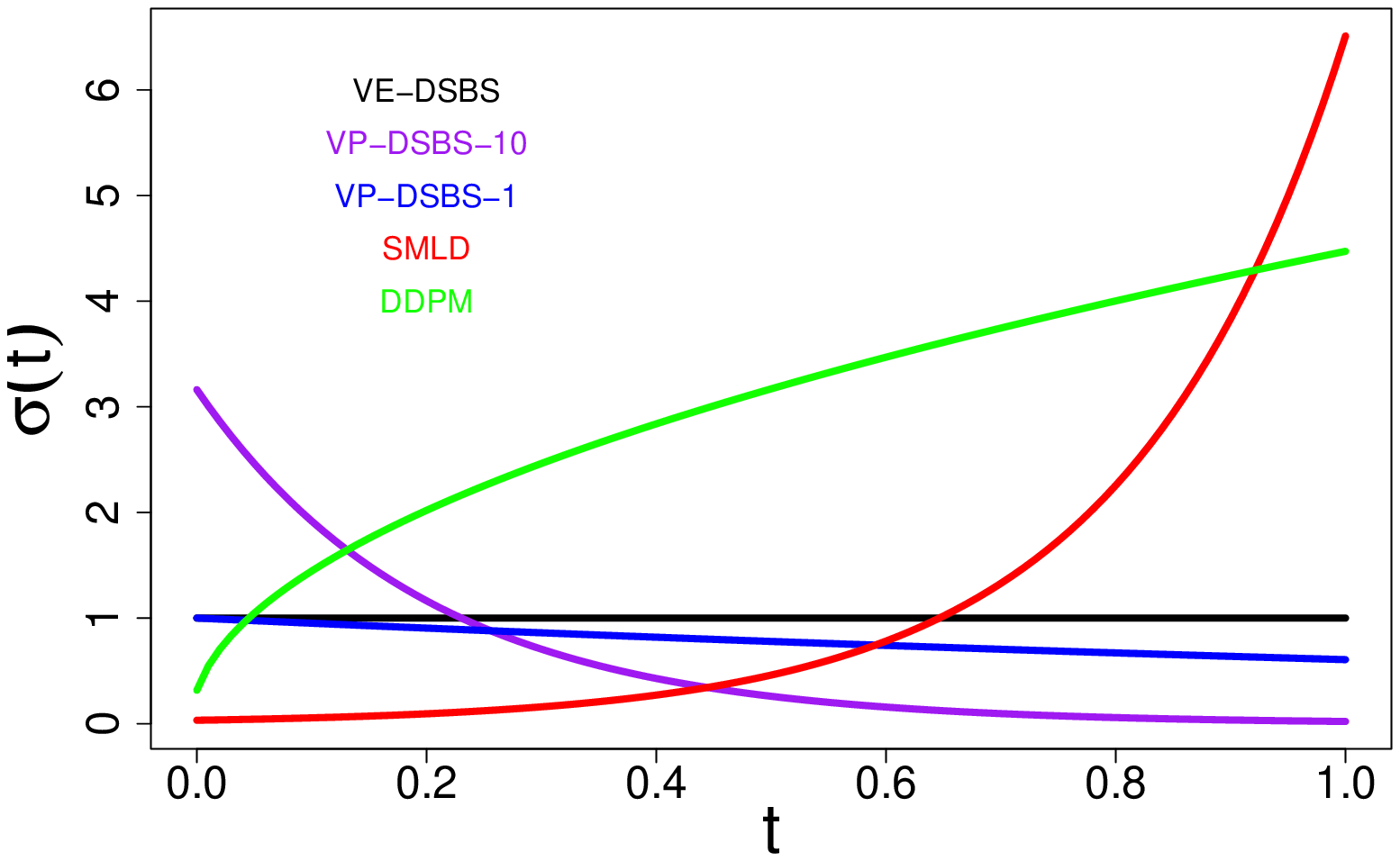,width=7.6cm,angle=-90}
	\end{center}
    \vspace{-0.5cm}
	\caption{Comparison of the diffusion function $\sigma(t)$ for different sampling schemes. VE-DSBS denotes data-based Schr\"{o}dinger Bridge sampler with VE SDE. VP-DSBS-1 and VP-DSBS-10 denote data-based Schr\"{o}dinger Bridge sampler with VP SDE (\ref{vptau}) for $\tau=1$ and $\tau=10$ respectively. SMLD denotes score matching with Langevin dynamics. DDPM denotes denoising diffusion probabilistic modeling. }
	\label{figure00}
\end{figure}

We have compared the numerical results of DSBS using three types of reference SDEs introduced in \ref{refsde} and found that the performance of sub-VP is worse than VE and VP in all the data sets to which we applied. Therefore, we only report the results for VE and VP here which are denoted by VE-DSBS and VP-DSBS respectively. For VE models, the noise scale is chosen to be $\sigma^2(t)=t$. The corresponding reference SDE process is just the standard Brownian motion, i.e. $d\bx_t=d\bw_t$, and its perturbation transition kernel can be derived as $q(0,\bx_0,t,\bx_t)=N(\bx_t;\bx_0,t\bI_d)$. Usually VE models normalize image inputs to the range $[0,1]$.  For VP models, we let $\beta(t)=\tau\exp(-\tau t)$ with $\tau=1$ or $\tau=10$. This corresponds to the following instantiation of the VP SDE:
\begin{eqnarray}\label{vptau}
d\bx_t=-\frac{1}{2}\tau\exp(-\tau t)\bx_t+\sqrt{\tau\exp(-\tau t)}d\bw_t,~t\in[0,1].
\end{eqnarray}
The perturbation kernel is given by 
\begin{eqnarray}\nn
q(0,\bx_0,t,\bx_t)=N(\bx_t;\bx_0e^{-\frac{1}{2}(1-e^{-\tau t})},[1-e^{-(1-e^{-\tau t})}]\bI_d)).
\end{eqnarray}
Figure \ref{figure0} compares the diffusion function $\sigma(t)$ used in various diffusion models including VE-DSBS, VP-DSBS with $\tau=1$, VP-DSBS with $\tau=10$, SMLD, and DDPM. The two score based generative models, SMLD and DDPM, involve sequentially corrupting data with slowly increasing noise, therefore their diffusion functions increase with $t$ as shown by the red and green curves in Figure \ref{figure00}. On the other hand, our SB based approaches reverse the step and thus form a sequence of decreasing noise scales. This is why their corresponding diffusion functions decrease with $t$ or keep to be a constant as shown by the purple, blue, and black curves in Figure \ref{figure00}.

In our experiment, we use the Euler-Maruyama discretization for the SDE (\ref{euler}) with a fixed step size, i.e. $t_j=j\delta,~j=0,1,\cdots,N, ~with~\delta=1/N$.

In addition to the visualization plot of data vs generated sample paths, we use some metrics described in Sections \ref{lowdim} and \ref{image} to evaluate the accuracy of our generators. 

\subsection{Learning low-dimensional distributions}\label{lowdim}

\begin{figure}[ht!]
    \vspace{-0.5cm}
		\begin{minipage}[t]{0.47\linewidth}
\centering
\includegraphics[width=1.08\textwidth]{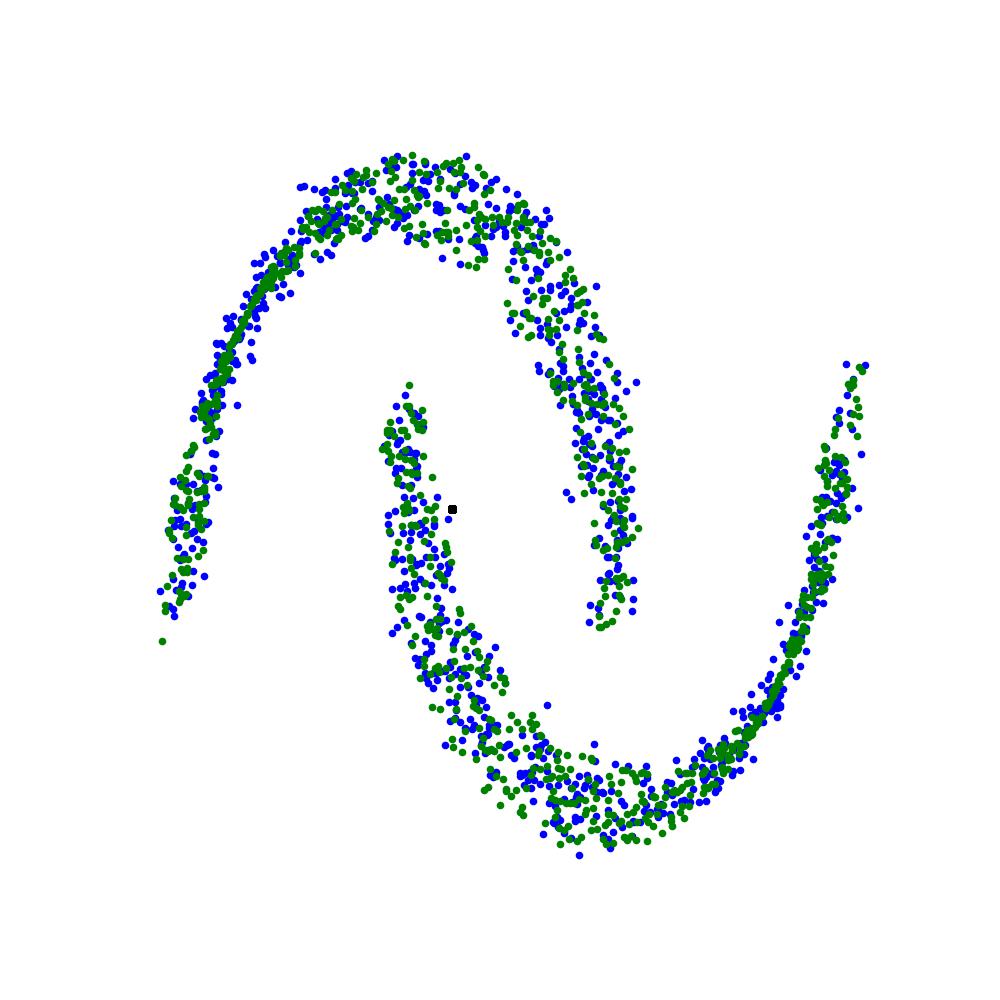}
		\end{minipage}
		\hspace{0.1cm}
		\begin{minipage}[t]{0.47\linewidth}
\centering
\includegraphics[width=1.08\textwidth]{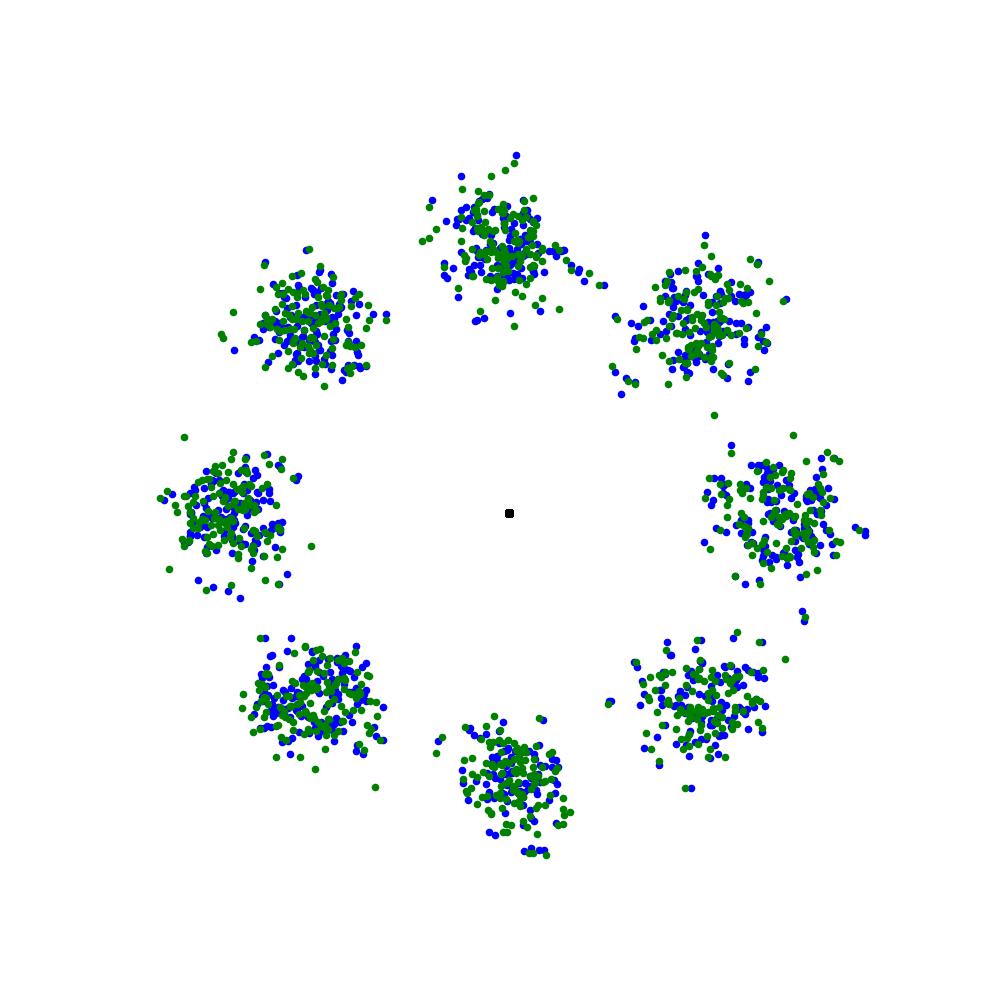}
		\end{minipage}
    \vspace{-0.5cm}
\caption{Visualizing the data of the Moons data set (left) and eight Gaussians data set (right). Green represents the training sample and blue represents the generated samples from our DSBS method.}
\label{figure0}
\end{figure}

We first evaluate how well various methods can effectively learn multi-modal distributions on low-dimensional data sets. We follow the setup from \cite{shi2023diffusion,tong2024} and use two common benchmark examples to illustrate and compare the different approaches. The first example is the Moons data which is sampled from a complicated distribution whose support is split into two disjoint regions of equal mass shaped like half-moon \citep{pedregosa11a}. The second example is 8-Gaussians data which is generated from a mixture of Gaussians with 8 components which are relatively far away from each other.  Similar to the methods and models reported in \cite{tong2024}, we use a training set size of 10000 and a test set size of 10000. We let the starting point $\bx_0=0$ and set the number of discretization steps $N=100$. We run each setting 10 times with mean and standard deviation reported. Figure \ref{figure0} compares the training samples (green) with the samples generated from our DSBS method (blue). They are very close and all the modes are included in both examples. 

We evaluate the empirical 2-Wasserstein distance for 10000 generated samples from our model. The value reported for ${\cal W}_2$ is 
\begin{eqnarray}\nn
{\cal W}_2&=&\left(\min_{\pi\in U(\hat{p}_1,q_1)}\int\|\bx-\by\|^2_2d\pi(\bx,\by)\right)^{1/2},
\end{eqnarray}
where $\hat{p}_1$ is sampled via our algorithm, $q_1$ is the test set, and $U(\hat{p}_1,q_1)$ is the set of all couplings of $\hat{p}_1$ and $q_1$. We report the 2-Wasserstein distance between the predicted distribution and the target distribution with test samples of size 10000. Table \ref{table1} summarizes the comparison of our approaches with the following methods:
\begin{itemize}
\item Simulation-Free Schr\"{o}dinger Bridges via Score and Flow Matching \citep{tong2024} with exact optimal transport couplings ($[SF]^2$M-exact) and with independent couplings ($[SF]^2$M-I);
\item Iterative Schr\"{o}dinger bridge models: diffusion Schr\"{o}dinger bridges (DSB) and diffusion Schr\"{o}dinger bridge matching with algorithms based on iterative proportional fitting (DSBM-IPF) and Iterative Markovian Fitting (DSBM-IMF);
\item ODE flow-based models: optimal transport conditional flow matching (OT-CFM), Schr\"{o}dinger bridges conditional flow matching (SB-CFM), independent conditional flow matching (I-CFM), rectified flow (RF), and flow matching  (FM).
\end{itemize}
The results derived from SDE based stochastic methods and ODE based deterministic methods are put in the top and bottom blocks in Table \ref{table1} respectively. They are reported in \cite{tong2024}. For 8-gaussians data set, VE-DSBS outperforms all methods and VP-DSBS-10 is in the third place. For Moons data set, VE-DSBS is in the fifth place. The performance of VE-DSBS is better than the performance of VP-DSBS-1 and VP-DSBS-10 on both data sets. The results from Table \ref{table1} show that DSBS is a competitive generative model for low-dimensional data. 
\begin{table}
  \begin{center}
    \caption{Summary of generative modelling performance for various methods on two-dimensional data in terms of ${\cal W}_2$. Mean and standard deviation based on 10 replication are reported.}
    \label{table1}
    \vspace{0.35cm}
    \begin{tabular}{c|c|c}\hline
      Algorithm&8-Gaussians&Moons\\\hline
      VE-DSBS&${\bf 0.267~(0.079)}$&0.148 (0.026)\\
      VP-DSBS-1&0.316 (0.062)&0.185 (0.020)\\
      VP-DSBS-10&0.293 (0.077)&0.329 (0.036)\\
      $[SF]^2$M-exact \citep{tong2024}&0.275 (0.058)&${\bf 0.124~(0.023)}$\\
      $[SF]^2$M-I \citep{tong2024}&0.393 (0.054)&0.185 (0.028)\\
      DSBM-IPF \citep{shi2023diffusion}&0.315 (0.079)&0.140 (0.006)\\
      DSBM-IMF \citep{shi2023diffusion}&0.338 (0.091)&0.144 (0.024)\\
      DSB \citep{debortoli2023diffusion}&0.411 (0.084)&0.190 (0.049)\\\hline
      OT-CFM \citep{tong2024improving}&0.303 (0.053)&0.130 (0.016)\\
      SB-CFM \citep{tong2024improving}&2.314 (2.112)&0.434 (0.594)\\
      RF \citep{liu2022rectified}&0.421 (0.071)&0.283 (0.045)\\
      I-CFM \citep{tong2024improving}&0.373 (0.103)&0.178 (0.014)\\
      FM \citep{lipman2023flow}&0.343 (0.058)&0.209 (0.055)\\\hline
    \end{tabular}
\end{center}
  \vspace{-0.25cm}
\end{table}

\subsection{Image Generation}\label{image}
\begin{figure}[ht!]
		\begin{minipage}[t]{0.49\linewidth}
\centering
\includegraphics[width=0.95\textwidth]{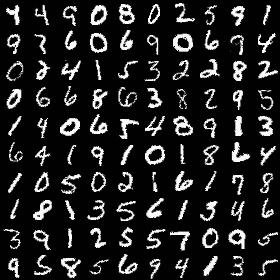}
		\end{minipage}
		\hspace{0.3cm}
		\begin{minipage}[t]{0.49\linewidth}
\centering
\includegraphics[width=0.95\textwidth]{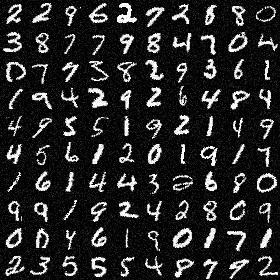}
		\end{minipage}
\caption{Comparison between true images (left) and generated images (right) on MNIST.}
\label{figure1}
\end{figure}

\begin{figure}[ht!]
		\begin{minipage}[t]{0.49\linewidth}
\centering
\includegraphics[width=0.95\textwidth]{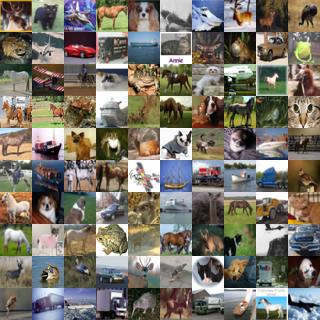}
		\end{minipage}
		\hspace{0.3cm}
		\begin{minipage}[t]{0.49\linewidth}
\centering
\includegraphics[width=0.95\textwidth]{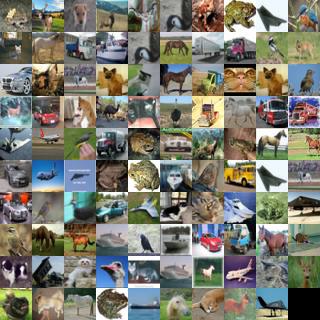}
		\end{minipage}
\caption{Comparison between true images (left) and generated images (right) on CIFAR10.}
\label{figure2}
\end{figure}

\begin{figure}[ht!]
\centering
\includegraphics[width=0.5\textwidth]{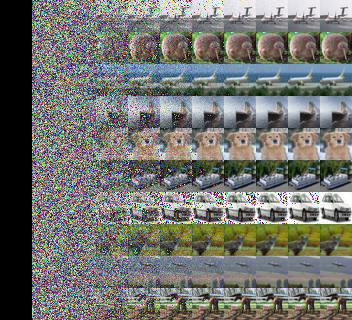}
\centering
\caption{Illustration of the CIFAR10 image generation process over time from left to right with intermediate samples.}
\label{figure3}
\end{figure}

Next, we validate our method on high-dimensional image generation. We consider the well-studied MNIST and CIFAR10 data sets. The MNIST data set consists of gray-valued digital images, each having 28$\times$28 pixels and showing one hand-written digit. The generated images for MNIST using the DSBS algorithm based on 10000 training samples are presented in Figure \ref{figure1}, which clearly suggest that our method is able to synthesize high-fidelity images. The CIFAR10 data set consists of 60000 images in 10 classes, with 6000 images per class. The images are colored and of size 32$\times$32 pixels. Figure \ref{figure2} shows the generated samples and their comparison with the true samples on CIFAR10. As the Figure demonstrates, our algorithm successfully generates sharp, high-quality, and diverse samples on high-dimensional image. Figure \ref{figure3} shows the progressive generation on CIFAR10 in our algorithm. It shows that our approach provides a valid path for the particles to move from a fixed point at $t=0$ to the target distribution at $t=1$. 

Table \ref{table2} summarizes the Fr\'{e}chet Inception Distance score (FID) \citep{heusel2018gans} on CIFAR10 of our algorithm as well as other two type of state-of-the-art generative models: Optimal Transport (OT) methods and score based generative models (SGM). The OT type of methods include SB-FBSDE, forward-backward SDE \citep{chen2023likelihood}, DOT, discriminator optimal transport \citep{tanaka2023discriminator}, multi-stage SB \citep{wang2021deep}, and DGflow, deep generative gradient flow \citep{ansari2021refining}. The SGM type of models include ScoreFlow, score-based generative flow \citep{song2021maximum},   VDM, variational diffusion models \citep{kingma2023variational}, and LSGM, latent score-based generative model \citep{vahdat2021scorebased}. The FID score is a metric used to assess the quality of images created by a generative model. It compares the distribution of generated images with the distribution of a set of real images. We report the FID over 30k samples w.r.t. the training set. Notably, our method achieves score 3.88 on CIFAR10, which is comparable to the top existing SGM type of methods and outperforms most of the other OT type of methods by a large margin in terms of the sample quality. We omit FID scores on MNIST as the scores on this data set are not widely reported. 

\begin{table}
  \begin{center}
    \caption{CIFAR10 evaluation using FID score w.r.t. the training set. }
      \vskip 0.1in
      \begin{tabular}{llcccccccc}
        \toprule
        {Model Class} & {Method}   & {FID } \\
        \midrule
        \multirow{4}{*}{{Optimal Transport }}
        & \textbf{VP-DSBS-10 (ours)}             &   3.88 \\[1pt]
        & SB-FBSDE \citep{chen2023likelihood}           &   3.01 \\[1pt]
        & DOT \citep{tanaka2023discriminator}  &   15.78 \\[1pt]
        & Multi-stage SB \citep{wang2021deep}       & 12.32 \\[1pt]
        & DGflow \citep{ansari2021refining}         &  9.63 \\[1pt]
        \midrule
        \multirow{5}{*}{{SGMs}}
        & SDE (deep, sub-VP; \citet{song2021scorebased})                   & 2.92  \\[1pt]
        & ScoreFlow \citep{song2021maximum}              & 5.7  \\[1pt]
        & VDM \citep{kingma2023variational}              & 4.00  \\[1pt]
        & LSGM \citep{vahdat2021scorebased}                           &\textbf{2.10} \\[1pt]
        \bottomrule
      \end{tabular} 
\label{table2}
 \end{center}
 \end{table}

\section{Conclusion}\label{conclusion}

In this article, we propose a novel class of simulation-free generative model for sampling from distributions which are known only through a finite set of samples. The procedure is through Euler-Maruyama discretization of a tractable class of Schr\"{o}dinger bridge diffusion process whose initial distribution is of degenerate type, i.e. the process starts from a fixed point. We obtain the solution of the diffusion process by deriving the explicit formulation for the drift function of its corresponding SDE. The formulation turns out to be quite simple and relies solely on data samples and the transition probability of the reference process. Our approach is more flexible than existing related works in that its reference process is not limited to standard Brownian motion but can be any Ito's process. We consider three type of reference SDEs: VE, VP, and sub-VP in our numerical studies and all of them have closed form expressions. The usefulness of our method is demonstrated through application to both low-dimensional simulated data and high-dimensional image data. The numerical results show that our method is competitive with the state-of-the-art generative models. The biggest advantage of our method is that it is simple, efficient, and low cost computationally as it is a one-step procedure and does not involve iteration and training the neural network.    

Our formulation opens up new algorithmic opportunities for developing efficient nonlinear diffusion models that have a broad scope of applicability. We plan to apply our method to more real-world data. We have considered three prior processes in our implementation and observed that the performance depends on the prior process and the optimal choice of the prior is different for different data sets. Many applications require more general priors whose distributions are not available in closed form if one considers SDEs with nonlinear drift functions. Future work can consider how to extend the present methodology to general situations and find the way on how to choose optimal prior process for given data samples.


\appendix

\section{Proof of Theorem \ref{thm}}\label{proof}

\begin{proof}

Assume that $\mQ$ is absolutely continuous with respect to the reference measure $\bP$, i.e. $D(\mQ\|\bP)\textless\infty$. The disintegration theorem then gives
\begin{eqnarray}\label{kld}
D(\mQ\|\bP)=D(\mQ_1\|\bP_1)+\int_{\bR^d}\mQ_1(d\bx)D(\mQ^{\bx}\|\bP^{\bx}),
\end{eqnarray}
where $\mQ^{\bx}$ (respectively, $\bP^{\bx}$) denotes the conditional probability law of $\bx_{t\in[0,1)}$ given $\bx_1=\bx$ under $\mQ$ (respectively, $\bP$). The conditional probability measure $\bP^{\bx}$ gives the probability law of the reference process which usually has a closed form solution. Now since $\mQ\in{\cal D}(\delta_{\ba},\mu)$, we have $\mQ_1=\mu$, whereas $\bP_1=\mu^{\bx}$ for the reference process $\bP$. Thus we have the following expression for the right hand side of (\ref{kld})
\begin{eqnarray}\nn
D(\mQ\|\bP)=D(\mu\|\mu^{\bx})+\int_{\bR^d}\mu(d\bx)D(\mQ^{\bx}\|\bP^{\bx})\ge D(\mu\|\mu^{\bx}),
\end{eqnarray}
where equality holds if and only if $\mQ^{\bx}=\bP^{\bx}$ almost everywhere. This immediately implies that $D(\mu\|\mu^{\bx})=\inf_{{\cal D}(\delta_{\ba},\mu)}D(\mQ\|\bP)$ and the above infimum is attained by the probability measure $\mQ^\bx(\cdot)=\int_{\bR^d}\mu(d\bx)\bP^{\bx}(\cdot)$, i.e. by the $\mu$-mixture of reference processes $\bP^{\bx}$.

Recall that the path measure $\bP$ describes a solution to the following SDE
\begin{eqnarray}\nn
d\bx_t=\bb(\bx_t,t)dt+\sigma(t) d\bw_t,~\bx_0=\ba.
\end{eqnarray}
We first make the gradient ansatz and assume that the diffusion process $\mQ$ is governed by an Ito SDE of the form 
\begin{eqnarray}\label{sde2}
d\bx_t=[\bb(\bx_t,t)-\sigma(t)^2{\nabla} \phi(\bx_t,t)]dt+\sigma(t) d\bw_t,~\bx_0=\ba,
\end{eqnarray}
where $\nabla$ denotes the gradient with respect to the space variable. Then we will show that we can choose a suitable function $\phi(\bx,t): \bR^d\times [0,1]\rightarrow\bR$ which is twice continuously differentiable in $\bx$ and once differentiable in $t$ such that the probability law of the resulting process $\bx_t$ defined in $[0,1]$ will be governed by $\mQ$. 

Given two Ito processes with the same diffusion, we can use the Girsanov theorem to write down the Radon-Nikodym derivative of $\mQ$ with respect to $\bP$ by
\begin{eqnarray}\label{rnd}
\frac{d\mQ}{d\bP}=\exp\left(-\int_0^1\frac{\sigma(t)^2}{2}\|{\nabla}\phi(\bx_t,t)\|^2dt-\int_0^1\sigma(t){\nabla}\phi(\bx_t,t)\cdot d\bw_t\right).
\end{eqnarray}
Let us define the process $(\phi_t)_{t\in[0,1]}$ by $\phi_t=\phi(\bx_t,t)$. Ito's lemma then gives 
\begin{eqnarray}\nn
d\phi(\bx_t,t)=\left(\partial_t\phi(\bx_t,t)+\nabla \phi(\bx_t,t)\cdot \bb(\bx_t,t)+\frac{\sigma(t)^2}{2}\bigtriangleup \phi(\bx_t,t)\right)dt+\sigma(t)\nabla \phi(\bx_t,t)\cdot d\bw_t.
\end{eqnarray}
where $\bigtriangleup=\nabla\cdot\nabla$ is the Laplacian. Integrating and rearranging, we obtain
\begin{eqnarray}\nn
&&-\int_0^1\sigma(t)\nabla \phi(\bx_t,t)\cdot d\bw_t\\\nn
&=&\int_0^1\left(\partial_t\phi(\bx_t,t)+\nabla \phi(\bx_t,t)\cdot\bb(\bx_t,t)+\frac{\sigma(t)^2}{2}\bigtriangleup \phi(\bx_t,t)\right)dt+\phi(\bx_0,0)-\phi(\bx_1,1).
\end{eqnarray} 
Substituting this into (\ref{rnd}) and using the definition of $\phi_t$ gives 
\begin{eqnarray}\nn
\frac{d\mQ}{d\bP}&=&\exp\left\{\phi(\bx_0,0)-\phi(\bx_1,1)\right.\\\nn
&&\left.+\int_0^1\left(\partial_t\phi(\bx_t,t)+\nabla \phi(\bx_t,t)\cdot\bb(\bx_t,t)+\frac{\sigma(t)^2}{2}\bigtriangleup \phi(\bx_t,t)-\frac{\sigma(t)^2}{2}\|{\boldsymbol\nabla} \phi(\bx_t,t)\|^2\right)dt\right\}.
\end{eqnarray}
Let $\phi(\bx,t)$ solves the PDE
\begin{eqnarray}\label{pde}
\partial_t\phi(\bx,t)+\nabla \phi(\bx,t)\cdot\bb(\bx,t)+\frac{\sigma(t)^2}{2}\bigtriangleup \phi(\bx,t)-\frac{\sigma(t)^2}{2}\|{\boldsymbol\nabla} \phi(\bx,t)\|^2=0
\end{eqnarray}
for all $(\bx,t)\in\bR^d\times[0,1]$ subject to the terminal condition $\phi(\bx_0,0)=0$ and $\phi(\bx,1)=-\log\frac{\mu}{\mu^\bx}(\bx)$. Now the PDE (\ref{pde}) is nonlinear in $\phi(\bx,t)$ due to the presence of the squared norm of the gradient of $\phi(\bx,t)$ on the right hand side. From the theory of PDE we can convert it into a linear PDE by making the logarithmic transformation $\phi(\bx,t)=-\log h(\bx,t)$ which is also called the Cole-Hopf transformation \citep{hopf1950}. Substituting this into (\ref{pde}), we obtain a much nicer linear PDE
for $h(\bx,t)$: 
\begin{eqnarray}\label{lpde}
\partial_th(\bx,t)+\nabla h(\bx,t)\cdot\bb(\bx_t,t)+\frac{\sigma(t)^2}{2}\bigtriangleup h(\bx,t)=0
\end{eqnarray}
on $\bR^d\times[0,1]$ subject to the terminal condition $h(\bx,0)=1$ and $h(\bx,1)=\frac{\mu}{\mu^\bx}(\bx)$. Let us again define the process $(h_t)_{t\in[0,1]}$ by $h_t=h(\bx_t,t)$ and apply Ito's lemma:
\begin{eqnarray}\nn
dh(\bx_t,t)&=&\partial_th(\bx_t,t)+\nabla h(\bx_t,t)\cdot\bb(\bx_t,t)+\frac{\sigma(t)^2}{2}\bigtriangleup h(\bx_t,t)+\sigma(t)\nabla h(\bx_t,t)\cdot d\bw_t\\\nn
&=&\sigma(t)\nabla h(\bx_t,t)\cdot d\bw_t,
\end{eqnarray}	 
where we have used the fact that $h(\bx,t)$ solves (\ref{lpde}). Taking integration we obtain 
\begin{eqnarray}\nn
h(\bx_1,1)-h(\bx_t,t)&=&\int_t^1\sigma(t)\nabla h(\bx_t,t)\cdot d\bw_t.
\end{eqnarray}	 
Taking conditional expectation on both side given $\bx_t=\bx$, we obtain 
\begin{eqnarray}\label{solution}
h(\bx,t)=\mE_{\bP}\left[\frac{\mu}{\mu^\bx}(\bx_1)|\bx_t=\bx\right]=\int g_t(\bx,\bx_1)\mu(d\bx_1),
\end{eqnarray}	 
for any $\bx\in\bR^d$ and $t\in[0,1]$, where $g_t(\bx,\bx_1)$ is defined in (\ref{transition}). This is also called the Feynman-Kac formula for the solution of (\ref{lpde}), one of the remarkable connections between the theory of PDE's and diffusion processes. To verify that the terminal conditions are satisfied, we obtain  
\begin{eqnarray}\nn
h(\bx_0,0)=\mE_{\bP}\left[\frac{\mu}{\mu^\bx}(\bx_1)|\bx_0=\bx_0\right]=\int\mu(\bx_1)d\bx_1=1.
\end{eqnarray}	 
Then transform (\ref{solution}) for $h(\bx,t)$ into $\phi(\bx,t)$ and substitute into (\ref{sde2}), we obtain 
\begin{eqnarray}\nn
\phi(\bx,t)&=&-\log\left[\int g_t(\bx,\bx_1)\mu(d\bx_1)\right],
\end{eqnarray}	 
which leads to  (\ref{drift0}), i.e.
\begin{eqnarray}\nn
\bu^\star(\bx,t)=-\sigma(t)^2\nabla\phi(\bx,t)=\frac{\sigma(t)^2\int\nabla_{\bx}g_t(\bx,\bx_1)\mu(d\bx_1)}{\int g_t(\bx,\bx_1)\mu(d\bx_1)}.
\end{eqnarray}	 

\end{proof}

\section{Gaussian mixture distributions}\label{gmm}

Assume that the target distribution $\mu$ is a Gaussian mixture, i.e.,
\begin{eqnarray}\nn
\mu=\sum_{k=1}^K\pi_k N(\btheta_k,\bSigma_k),~\sum_{k=1}^K\pi_k=1,~k=1,\cdots,K,
\end{eqnarray}	 
where $K$ is the number of mixture components. For VP SDE, the drift coefficient (\ref{vp}) can be written as
\begin{eqnarray}\nn
\bu^\star(\bx,t)&=&\kappa_1\frac{\sum_{k=1}^K\pi_k\rho_k\tilde{\btheta}_k}{\sum_{k=1}^K\pi_k\rho_k}-\kappa_2\bx,
\end{eqnarray}	 
where
\begin{eqnarray}\nn
\kappa_1&=&\frac{\beta(t)e^{-\frac{1}{2}\int_t^1\beta(s^\prime)ds^\prime}}{1-e^{-\int_t^1\beta(s^\prime)ds^\prime}},\\\nn
\kappa_2&=&\frac{\beta(t)e^{-\int_t^1\beta(s^\prime)ds^\prime}}{1-e^{-\int_t^1\beta(s^\prime)ds^\prime}},\\\nn
\tilde{\btheta}_k&=&\tilde{\bSigma}_k\left(\bSigma_k^{-1}\btheta_k+\frac{\bm_1}{\zeta_1}-\frac{\bm_2}{\zeta_2}\right),\\\nn
\rho_k&=&\left|\tilde{\bSigma}_k\right|^{1/2}|\bSigma_k|^{-1/2}\exp\left\{\frac{1}{2}\left[\tilde{\btheta}_k^T\tilde{\bSigma}_k^{-1}\tilde{\btheta}_k-\btheta_k^T\bSigma_k^{-1}\btheta_k\right]\right\},
\end{eqnarray}	 
where
\begin{eqnarray}\nn
\bm_1&=&\ba e^{-\frac{1}{2}\int_0^1\beta(s^\prime)ds^\prime},\\\nn
\bm_2&=&\bx e^{-\frac{1}{2}\int_t^1\beta(s^\prime)ds^\prime},\\\nn
\zeta_1&=&1-e^{-\int_0^1\beta(s^\prime)ds^\prime},\\\nn
\zeta_2&=&1-e^{-\int_t^1\beta(s^\prime)ds^\prime},\\\nn
\tilde{\bSigma}_k&=&\left\{\bSigma_k^{-1}+\left(\frac{1}{\zeta_1}-\frac{1}{\zeta_2}\right)\bI_d\right\}^{-1}.
\end{eqnarray}	 

\bibliographystyle{apalike} 
\bibliography{biblist}

\end{document}